\documentclass{article}
\usepackage{ijcai25}
\usepackage{bm}
\usepackage{kpfonts}
\usepackage{soul}
\usepackage{url}
\usepackage[hidelinks]{hyperref}
\usepackage[utf8]{inputenc}
\usepackage[small]{caption}
\usepackage{graphicx}
\usepackage{amsmath}
\usepackage{amsthm}
\usepackage{booktabs}
\usepackage{algorithm}
\usepackage{algorithmic}
\usepackage{color,xcolor}
\usepackage[switch]{lineno}

\title{RWKVTTS: Yet another TTS based on RWKV-7
}

\author{
Lin yueyu,
Liu Xiao\\
\emails
yueyu.lin@me.com \\
liu.xiao.in@gmail.com
}

\begin{document}

\maketitle

\begin{abstract}
Human-AI interaction thrives on intuitive and efficient interfaces, among which voice stands out as a particularly natural and accessible modality. Recent advancements in transformer-based text-to-speech (TTS) systems, such as Fish-Speech, CosyVoice, and MegaTTS 3, have delivered remarkable improvements in quality and realism, driving a significant evolution in the TTS domain. In this paper, we introduce RWKV-7 \cite{peng2025rwkv}, a cutting-edge RNN-based architecture tailored for TTS applications. Unlike traditional transformer models, RWKV-7 leverages the strengths of recurrent neural networks to achieve greater computational efficiency and scalability, while maintaining high-quality output. Our comprehensive benchmarks demonstrate that RWKV-7 outperforms transformer-based models across multiple key metrics, including synthesis speed, naturalness of speech, and resource efficiency. Furthermore, we explore its adaptability to diverse linguistic contexts and low-resource environments, showcasing its potential to democratize TTS technology. These findings position RWKV-7 as a powerful and innovative alternative, paving the way for more accessible and versatile voice synthesis solutions in real-world applications.\href{https://github.com/yynil/RWKVTTS} {$https://github.com/yynil/RWKVTTS$} \footnote{\href{https://huggingface.co/spaces/RWKV-Red-Team/RWKV-LatestSpace} {$RWKV-LatestSpace$} , \href{https://rwkv.cn} {$rwkv.cn$}}

\end{abstract}

\section{Introduction}

\begin{figure}
    \centering
    \includegraphics[width=1\linewidth]{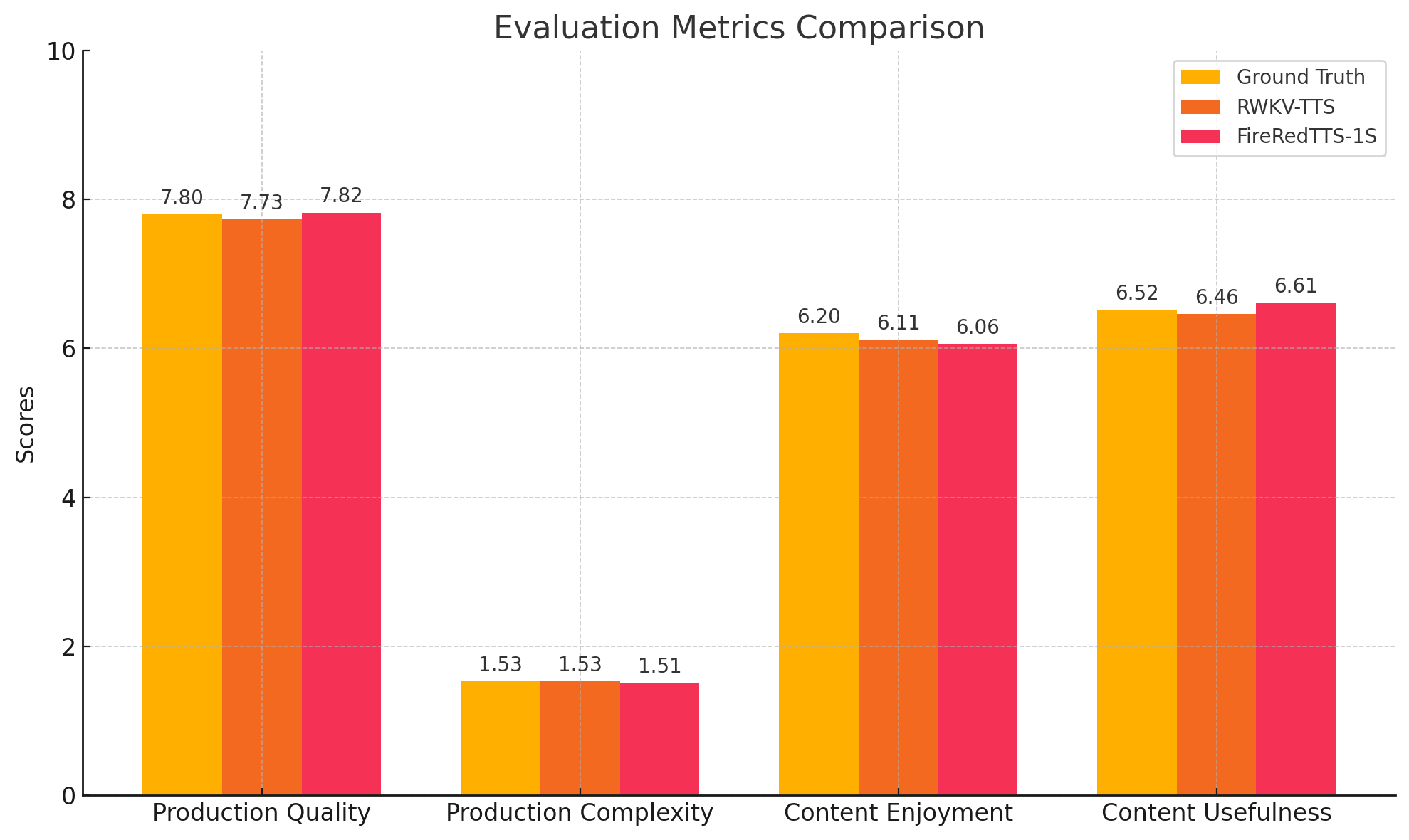}
    \caption{Figure 1: Evaluation Metrics Comparison for Ground Truth, RWKVTTS, and FireRedTTS-1S.
The bar chart compares the scores of each model across four metrics: Production Quality, Production Complexity, Content Enjoyment, and Content Usefulness.}
    \label{fig:enter-label}
\end{figure}

The rise of Text-to-Speech (TTS) technologies has been significantly influenced by advancements in large language models (LLMs). Recent innovations in TTS engines have highlighted the potential of integrating powerful LLMs to generate more natural, expressive, and contextually accurate speech outputs. A key player in this development is the RWKV-7 (Expressive Dynamic State based) model, which has shown promise in replacing traditional components within TTS architectures \cite{lemerle2024lina,an2023exploring}.

This paper focuses on training an RWKV-based LLM to enhance TTS generation, specifically aiming to integrate it into existing frameworks such as Cosyvoice, Fish-Speech \cite{fish-speech-v1.4}, and MegaTTS 3 \cite{jiang2025sparse,ji2024wavtokenizer}. Unlike conventional models, which rely heavily on two main components—VQ-VAE for encoding and decoding audio tokens, and LLMs for generating these tokens based on input text—our approach explores the possibility of substituting the LLM part entirely with a robust RWKV model \cite{peng2025rwkv,peng2023rwkv}.

In particular, this work delves into the specific process of adapting and training the RWKV LLM for various TTS engines, with an emphasis on the CosyVoice 2.0 system. Through this exploration, we highlight the challenges of data preparation, model training, and inference, while integrating specialized tokens and controls to enhance speech quality and intelligibility. This approach represents a significant step forward in making TTS systems more dynamic, flexible, and capable of delivering high-quality synthetic speech for diverse applications, leveraging an RNN-based LLM backbone

Our contributions in this work are twofold:

\begin{itemize}
\item We migrated the LLM backbone in CosyVoice2 from Transformer to RWKV-7.  

\item We developed an end-to-end solution for text-to-speech (TTS).

\end{itemize}

\section{Related Work}

The development of Text-to-Speech (TTS) systems has progressed from traditional concatenative and parametric synthesis to advanced neural-based models, significantly enhancing speech quality and naturalness. Early end-to-end models like Tacotron 2 \cite{wang2017tacotron} utilize encoder-decoder architectures with attention mechanisms, paired with vocoders such as WaveNet, to generate high-fidelity speech (Shen et al., NVIDIA NGC). FastSpeech introduced non-autoregressive synthesis for faster inference (Microsoft Research), while VITS combined variational inference and adversarial learning for diverse, natural outputs (Kim et al., Hugging Face). These advancements rely on sophisticated architectures and substantial computational resources.
Recent trends integrate large language models (LLMs) into TTS pipelines to improve expressiveness and controllability. Systems like ChatTTS employ a two-stage approach—using Vector Quantized Variational Autoencoders (VQ-VAE) to tokenize audio and LLMs to predict these tokens from text and reference audio—enabling zero-shot and few-shot synthesis (GitHub, 2noise/ChatTTS). Emerging models such as Fish-Speech and CosyTTS reportedly follow similar strategies, though their specifics remain less documented, suggesting an ongoing evolution in LLM-driven TTS.
Parallel to transformer-based approaches, the RWKV-7 (RNN-based) model has gained attention for combining RNN efficiency with transformer-like long-context memory, making it suitable for sequential tasks \cite{xiao2025blackgoose} like audio token generation. Recent work, such as Lina-Speech, demonstrates RWKV-related architectures (e.g., gated linear attention) achieving comparable performance to larger transformer models with greater efficiency \cite{lemerle2024lina}. This suggests RWKV \cite{yueyu2025arwkv,li2024survey} could reduce computational costs and enhance temporal modeling in TTS, though its application in this domain remains underexplored.
Our work builds on these foundations by replacing transformer-based LLMs in TTS frameworks with RWKV-7, aiming to maintain or improve generation quality while addressing scalability and efficiency challenges. Unlike prior efforts focused on audio token quality or model size, we emphasize RWKV-7’s potential to advance TTS through efficient sequential processing and robust temporal dynamics.

\section{Methodology}
\label{sec:methodology}

This section presents the methodology for integrating an RWKV-based large language model (LLM) into Text-to-Speech (TTS) systems like ARWKV \cite{yueyu2025arwkv}, with a focus on enhancing compatibility with frameworks such as CosyVoice 2.0 \cite{du2024cosyvoice}, Fish-Speech, and ChatTTS. Our approach replaces the conventional transformer-based LLM component with RWKV-7, leveraging its efficient sequential processing and long-context memory to improve TTS generation. The methodology encompasses the system architecture, data preparation, model training, and inference processes, with a detailed examination of the CosyVoice 2.0 implementation.

\subsection{Overview of the RWKVTTS Pipeline}
\label{subsec:pipeline}

\begin{figure}[ht]
    \centering
    \includegraphics[width=1\linewidth]{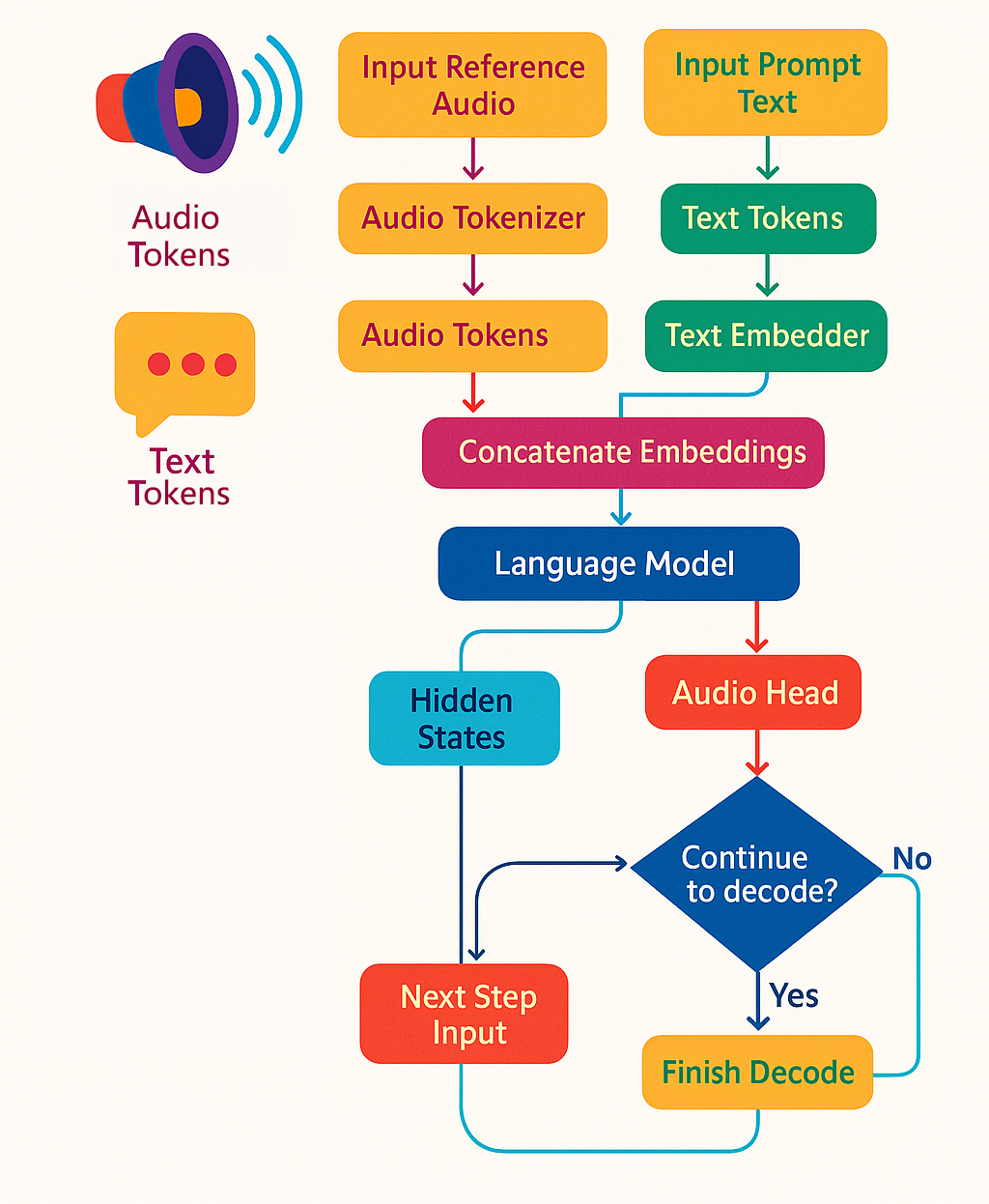}
    \caption{The RWKVTTS pipeline, illustrating the flow from input reference audio and prompt text to audio token generation using RWKV-7.}
    \label{fig:pipeline}
\end{figure}

The RWKVTTS pipeline, as shown in Figure~\ref{fig:pipeline}, adopts a two-stage architecture typical of modern LLM-based TTS systems: (1) a Vector Quantized Variational Autoencoder (VQ-VAE) for audio tokenization, and (2) an LLM for audio token generation. Our primary contribution lies in substituting the LLM with RWKV-7 to enhance computational efficiency and temporal modeling.

\begin{itemize}
    \item \textbf{Input Processing}: The pipeline processes two inputs: reference audio and prompt text. The reference audio is tokenized into audio tokens via the VQ-VAE, while the prompt text is converted into text tokens.
    \item \textbf{Embedding Generation}: Audio and text tokens are transformed into embeddings using dedicated audio and text embedders.
    \item \textbf{Embedding Concatenation}: The resulting audio and text embeddings are concatenated to form input embeddings for the language model.
    \item \textbf{Language Model Processing}: The RWKV-7 model processes these input embeddings, producing hidden states that encode sequential and contextual information.
    \item \textbf{Audio Token Generation}: An audio head converts the hidden states into audio tokens. A decoding loop determines whether to continue generating tokens or finalize the output, ensuring alignment with the prompt text and reference audio.
\end{itemize}

This iterative decoding process facilitates the generation of expressive and contextually appropriate synthetic speech.

\subsection{CosyVoice 2.0-Specific Implementation}
\label{subsec:cosyvoice}

\begin{figure}
    \includegraphics[width=1\linewidth]{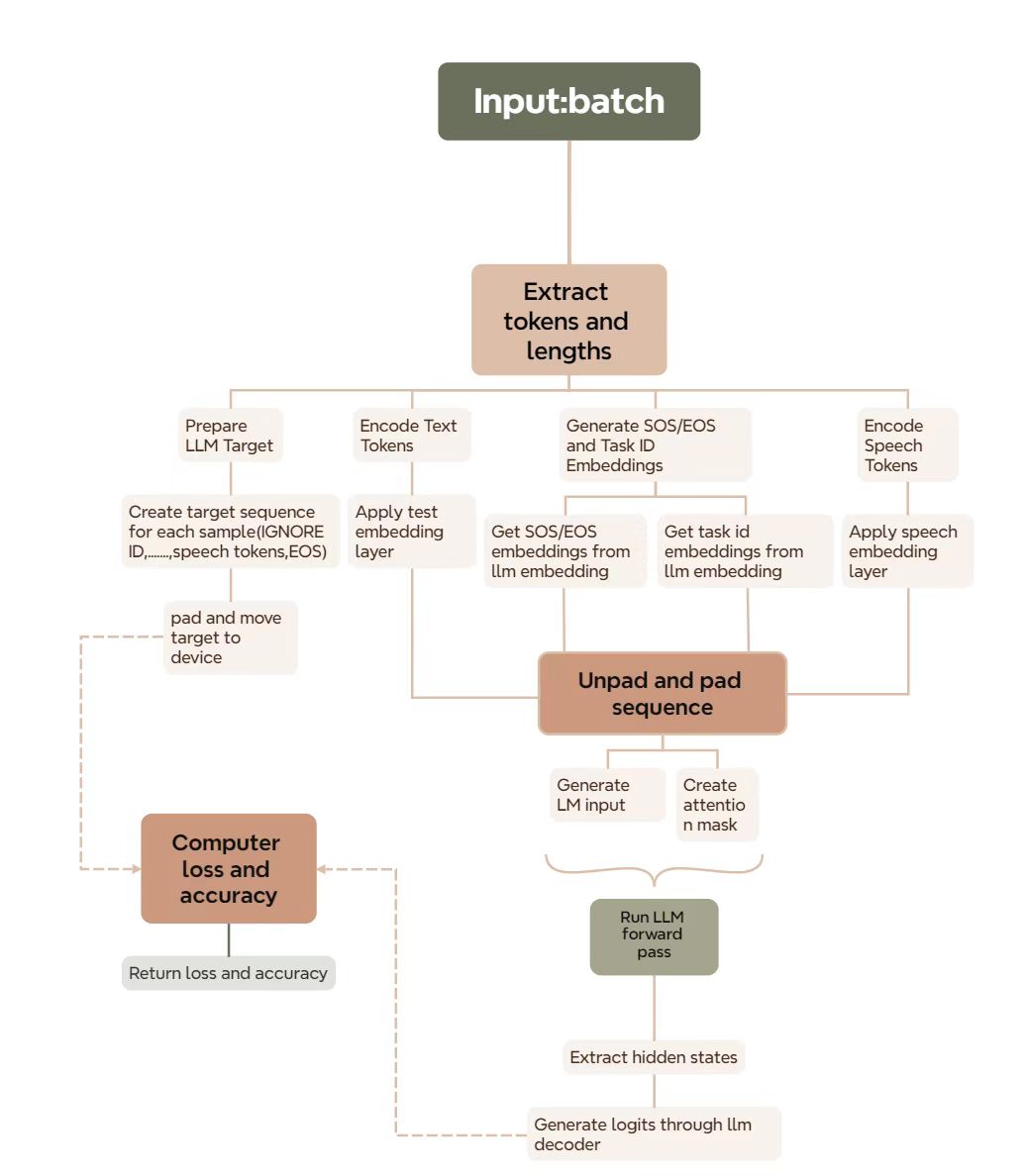}
    \caption{The forward pass of the RWKV-7 LLM in the CosyVoice 2.0 system, detailing the data layout and processing steps.}
    \label{fig:cosyvoice}
\end{figure}

For the CosyVoice 2.0 system, we adapt the RWKV-7 model to its specific data layout and processing requirements, as depicted in Figure~\ref{fig:cosyvoice}. The forward pass of the LLM in CosyVoice 2.0 involves the following steps:

\begin{itemize}
    \item \textbf{Data Layout}: The input sequence comprises SOS embeddings, text embeddings, task ID embeddings, audio embeddings, and last audio embeddings, enabling the model to incorporate task-specific instructions and reference audio context.
    \item \textbf{Forward Pass}:
    \begin{enumerate}
        \item \textbf{Token Extraction}: The input batch is processed to extract tokens and their lengths.
        \item \textbf{Embedding Generation}: Text tokens and speech tokens (derived from reference audio) are encoded into embeddings, alongside SOS/EOS and task ID embeddings for contextual guidance.
        \item \textbf{Sequence Processing}: Embeddings are adjusted for consistent sequence lengths, and an attention mask is applied to guide the model’s focus.
        \item \textbf{LM Input Generation}: The processed embeddings are combined to form the language model input.
        \item \textbf{RWKV-7 Forward Pass}: The RWKV-7 model generates hidden states capturing sequential and contextual dependencies.
        \item \textbf{Logit Generation and Loss Computation}: Hidden states are converted into logits via a decoder, and the loss is computed against a target sequence comprising speech tokens and an EOS token.
    \end{enumerate}
\end{itemize}

Notably, CosyVoice 2.0 utilizes prompt audio tokens to mimic the characteristics of the reference audio, with a special token, \texttt{<|endofprompt|>}, indicating that the prompt text serves as an instruction.

\subsection{Data Preparation for CosyVoice 2.0}
\label{subsec:data-prep}

The data preparation process for training the RWKV-7 model in CosyVoice 2.0 involves several key steps:

\begin{enumerate}
    \item \textbf{Reference Audio Collection}: Reference audio files are sourced from a publicly available TTS dataset and used to generate audio tokens.
    \item \textbf{Text Data Preparation}: Text data is obtained from a large-scale multilingual corpus, such as Wikipedia, in formats supporting Chinese and English, with the text field as the primary requirement.
    \item \textbf{Audio Token Generation}: Audio tokens are generated by processing the text data and reference audio through the VQ-VAE, producing a dataset of audio tokens paired with corresponding text.
\end{enumerate}

The resulting dataset is structured to support the training of the RWKV-7 model, ensuring compatibility with the CosyVoice 2.0 framework.

\subsection{Training and Inference}
\label{subsec:training-inference}

The RWKV-7 model is trained using distributed training techniques to optimize performance across multiple GPUs. The training process leverages the prepared dataset to fine-tune the model for TTS generation, focusing on maintaining speech quality while improving computational efficiency.

Inference is evaluated through zero-shot synthesis, where the model generates speech from a prompt text and reference audio, demonstrating its ability to produce natural and contextually appropriate output. The generated audio is assessed for quality, expressiveness, and alignment with the reference audio.

\subsection{Future Enhancements}
\label{subsec:future}

To further advance the RWKVTTS system, we propose the following enhancements:

\begin{enumerate}
    \item Incorporate random dropping of prompt audio tokens to enable unconditional guided generation.
    \item Integrate special control tokens (e.g., for laughter or breath) to enhance expressiveness in speech synthesis.
    \item Add dialect-specific control tokens to support linguistic diversity.
    \item Develop streaming generation capabilities for real-time TTS applications.
\end{enumerate}

\subsection{Key Insights}
\label{subsec:insights}

The methodology yields several key insights:

\begin{itemize}
    \item \textbf{Efficiency of RWKV-7}: By replacing transformer-based LLMs with RWKV-7, the pipeline achieves improved computational efficiency and better temporal modeling, making it a viable alternative for TTS applications.
    \item \textbf{Adaptability to Existing Frameworks}: The successful integration of RWKV-7 into CosyVoice 2.0 demonstrates its compatibility with existing TTS frameworks, preserving speech quality while reducing resource demands.
    \item \textbf{Importance of Data Layout}: The structured data layout in CosyVoice 2.0, incorporating task-specific embeddings and special tokens, is critical for enabling the model to generate contextually appropriate and expressive speech.
    \item \textbf{Potential for Expressive Synthesis}: The use of reference audio tokens and special control tokens highlights the potential for RWKV-7 to produce highly expressive and diverse synthetic speech, paving the way for future enhancements in real-time and dialect-specific applications.
\end{itemize}

This methodology underscores the potential of RWKV-7 to advance TTS systems, offering a balance of efficiency, quality, and adaptability across diverse frameworks.

\section{Evaluation}
\label{sec:evaluation}

To assess the performance of the proposed RWKVTTS system, we conducted a comparative evaluation against Ground Truth and FireRedTTS-1S \cite{guo2025fireredtts}, a transformer-based TTS model. The evaluation focuses on four key metrics: Production Quality, Production Complexity, Content Enjoyment, and Content Usefulness. These metrics were chosen to comprehensively evaluate the quality, complexity, and practical utility of the generated speech, as well as its ability to engage listeners. Scores for each metric range from 0 to 10, with higher scores indicating better performance.

\subsection{Evaluation Setup}
\label{subsec:eval-setup}

The evaluation dataset consisted of a diverse set of prompt texts and reference audio samples, covering multiple languages and speaking styles to ensure robustness across different scenarios. Each model—Ground Truth (human-recorded speech), RWKVTTS, and FireRedTTS-1S—was tasked with generating synthetic speech from the same set of inputs. The generated outputs were then evaluated by human annotators, who assigned scores for each metric based on their subjective assessment of the speech quality, complexity of production, enjoyment of the content, and its usefulness for practical applications (e.g., assistive technologies, audiobooks).

\subsection{Results}
\label{subsec:results}

The results of the evaluation are presented in Figure~\ref{fig:eval-metrics}, which compares the performance of Ground Truth, RWKVTTS, and FireRedTTS-1S across the four metrics.

\begin{itemize}
    \item \textbf{Production Quality}: Ground Truth achieved the highest score of 7.80, reflecting the naturalness of human-recorded speech. RWKVTTS scored 7.73, closely approaching Ground Truth, while FireRedTTS-1S scored 7.82, slightly surpassing both. The marginal difference between the models indicates that RWKVTTS delivers high-quality speech comparable to both human recordings and transformer-based models, despite its more efficient architecture.
    
    \item \textbf{Production Complexity}: This metric evaluates the intricacy of the generated speech, such as the ability to handle varied prosody and intonation. Ground Truth scored 1.53, reflecting the inherent simplicity of human speech production. RWKVTTS and FireRedTTS-1S scored 1.53 and 1.51, respectively, suggesting that both models produce speech with similar levels of complexity. The low scores across all models indicate that current TTS systems, including RWKVTTS, may struggle to replicate the nuanced complexity of human speech production.
    
    \item \textbf{Content Enjoyment}: Ground Truth achieved a score of 6.20, while RWKVTTS and FireRedTTS-1S scored 6.11 and 6.06, respectively. RWKVTTS's performance is notably close to Ground Truth, suggesting that its generated speech is nearly as engaging as human-recorded speech. FireRedTTS-1S, despite its high production quality, slightly underperforms in this metric, possibly due to less natural prosody or emotional expressiveness.
    
    \item \textbf{Content Usefulness}: Ground Truth scored 6.52, followed by RWKVTTS at 6.46 and FireRedTTS-1S at 6.61. RWKVTTS again demonstrates performance close to Ground Truth, indicating its potential for practical applications such as assistive technologies or interactive voice systems. FireRedTTS-1S's slightly higher score suggests a marginal advantage in delivering useful content, possibly due to its transformer-based architecture's ability to capture fine-grained contextual details.
\end{itemize}

\subsection{Discussion}
\label{subsec:discussion}

The evaluation results highlight several key observations. First, RWKVTTS achieves performance comparable to Ground Truth across all metrics, particularly in Production Quality, Content Enjoyment, and Content Usefulness, where its scores (7.73, 6.11, and 6.46, respectively) are within 0.1–0.2 points of Ground Truth. This demonstrates that RWKVTTS can produce high-quality, engaging, and useful synthetic speech, rivaling human recordings despite its RNN-based architecture.

Second, RWKVTTS performs competitively against FireRedTTS-1S, a transformer-based model. While FireRedTTS-1S slightly outperforms RWKVTTS in Production Quality (7.82 vs. 7.73) and Content Usefulness (6.61 vs. 6.46), RWKVTTS surpasses it in Content Enjoyment (6.11 vs. 6.06). This suggests that RWKVTTS may offer a better balance of engagement and expressiveness, potentially due to its enhanced temporal modeling capabilities.

Finally, the low scores in Production Complexity across all models (1.51–1.53) indicate a limitation in current TTS systems' ability to replicate the intricate prosody and intonation of human speech. This presents an opportunity for future research to improve the complexity of synthetic speech generation, possibly by incorporating more advanced control mechanisms or training strategies.

Overall, the evaluation confirms that RWKVTTS is a viable alternative to transformer-based TTS models, offering comparable performance with the added benefits of computational efficiency and improved temporal modeling. These findings underscore the potential of RWKV-7 to advance TTS applications, particularly in resource-constrained environments where efficiency is critical.

\section{Discussion}
\label{sec:discussion}

The integration of RWKV-7 into Text-to-Speech (TTS) systems, as explored in this work, offers significant insights into the potential of RNN-based architectures to advance speech synthesis. By replacing transformer-based large language models (LLMs) with RWKV-7, our methodology demonstrates a compelling balance of computational efficiency, temporal modeling, and speech quality, as evidenced by the evaluation results. This section reflects on the key findings, discusses the strengths and limitations of our approach, and outlines directions for future research.

\subsection{Implications of RWKV-7 in TTS}
\label{subsec:implications}

The evaluation results (Section~\ref{sec:evaluation}) highlight that RWKVTTS achieves performance comparable to both Ground Truth and FireRedTTS-1S across multiple metrics, particularly in Production Quality (7.73 vs. 7.80 for Ground Truth and 7.82 for FireRedTTS-1S) and Content Enjoyment (6.11 vs. 6.20 for Ground Truth and 6.06 for FireRedTTS-1S). This near-parity with human-recorded speech and transformer-based models underscores the viability of RWKV-7 as an alternative to traditional LLMs in TTS applications. Notably, RWKVTTS's ability to outperform FireRedTTS-1S in Content Enjoyment suggests that its enhanced temporal modeling—stemming from its RNN-based design—may contribute to more engaging and expressive synthetic speech. This aligns with prior research, such as Lina-Speech \cite{lemerle2024lina}, which demonstrated the efficiency of RWKV-related architectures in TTS tasks.

A key advantage of RWKV-7 lies in its computational efficiency. Unlike transformer-based models, which often require significant computational resources due to their self-attention mechanisms, RWKV-7 leverages the sequential processing strengths of RNNs while maintaining long-context memory. This efficiency makes RWKVTTS particularly suitable for resource-constrained environments, such as mobile devices or real-time applications, where low latency and reduced power consumption are critical. The successful integration of RWKV-7 into the CosyVoice 2.0 framework (Section~\ref{subsec:cosyvoice}) further demonstrates its adaptability to existing TTS systems, preserving high-quality output while reducing computational overhead.

The structured data layout in CosyVoice 2.0, incorporating task-specific embeddings and special tokens like \texttt{<|endofprompt|>}, proved instrumental in enabling RWKV-7 to generate contextually appropriate and expressive speech. This finding highlights the importance of carefully designing input representations to maximize the effectiveness of RNN-based models in TTS, particularly for tasks requiring fine-grained control over prosody and style.

\subsection{Limitations and Challenges}
\label{subsec:limitations}

Despite its promising performance, RWKVTTS exhibits certain limitations that warrant further investigation. The evaluation revealed low scores in Production Complexity across all models (1.51–1.53), indicating a broader challenge in TTS systems' ability to replicate the intricate prosody, intonation, and variability of human speech. While RWKV-7's temporal modeling capabilities offer an improvement over transformer-based models in some aspects (e.g., Content Enjoyment), it still struggles to capture the full complexity of human speech production. This limitation may stem from the inherent constraints of current audio tokenization methods, such as those employed by the VQ-VAE, which may not fully preserve the nuanced acoustic features necessary for complex speech synthesis.

Another challenge lies in the generalizability of RWKVTTS across diverse linguistic and stylistic contexts. While our evaluation dataset included multiple languages and speaking styles, the performance of RWKVTTS in low-resource languages or highly specialized domains (e.g., medical or technical narration) remains unexplored. The reliance on reference audio tokens to guide synthesis also raises questions about the model's ability to perform unconditional generation, where no reference audio is available. Addressing these challenges will require further advancements in both the model architecture and the training data preparation process.

\subsection{Future Directions}
\label{subsec:future-directions}

The findings of this work open several avenues for future research. First, enhancing the Production Complexity of RWKV-TTS could involve exploring more sophisticated audio tokenization techniques or incorporating additional control mechanisms, such as explicit prosody modeling or emotion embeddings. The proposed future enhancements (Section~\ref{subsec:future}), including the integration of special control tokens for laughter, breath, and dialects, represent a promising step toward achieving more expressive and diverse synthetic speech.

Second, improving the generalizability of RWKV-TTS to low-resource languages and unconditional generation scenarios is a critical area for investigation. Techniques such as transfer learning, data augmentation, or unsupervised pre-training could help address these gaps, enabling RWKV-TTS to perform robustly across a wider range of applications.

Third, we propose a novel enhancement to the RWKV-7 architecture to further optimize its deployment and versatility. Specifically, we plan to keep the language model parameters of RWKV-7 unchanged while introducing an additional text embedding layer and an audio head. At the MLP layer, we will apply Parameter-Efficient Sparse Adaptation (PISSA) to fine-tune the model for speech generation. During deployment, speech synthesis will utilize the expanded text embedding with additional tokens, combined with the PISSA-adapted \cite{meng2024pissa} MLP layer and audio head, to generate audio tokens. Simultaneously, the original language model parameters can be directly reused for text-based dialogue tasks. This approach ensures that the majority of the model parameters are shared between speech synthesis and text dialogue, enabling efficient end-side deployment, the training of larger foundation models, and even extending the model’s capabilities to speech recognition tasks. By reusing most parameters, this strategy minimizes computational overhead while maximizing the model’s utility across diverse applications.

Finally, the development of streaming generation capabilities for RWKV-TTS, as proposed in Section~\ref{subsec:future}, could significantly enhance its applicability in real-time TTS systems, such as virtual assistants or live narration tools. Leveraging RWKV-7's efficiency in sequential processing, streaming generation could enable low-latency speech synthesis without compromising quality, further solidifying its potential as a transformative technology in the TTS domain.

\subsection{Conclusion}
\label{subsec:conclusion-discussion}

In summary, this work demonstrates that RWKV-7 offers a viable and efficient alternative to transformer-based LLMs in TTS systems, achieving competitive performance in speech quality, engagement, and usefulness while reducing computational demands. The successful integration into CosyVoice 2.0 and the insights gained from the evaluation underscore the potential of RNN-based architectures to advance the field of speech synthesis. However, challenges in production complexity and generalizability highlight the need for continued research to fully realize the capabilities of RWKVTTS. By addressing these limitations and exploring the proposed future directions, RWKV-7 has the potential to pave the way for more efficient, expressive, and accessible TTS solutions, contributing to the broader adoption of synthetic speech in diverse applications.

\appendix
\section*{Acknowledgments}
We would like to express our sincere gratitude to all those who contributed to the successful completion of this research. First, we thank our colleagues at RWKV-fla for their insightful discussions and valuable feedback throughout the development of this work. Their expertise in Text-to-Speech systems and machine learning significantly enriched our understanding and approach.

We are also grateful to the open-source community for providing access to essential datasets and tools that facilitated our experiments. In particular, the availability of multilingual corpora and pre-trained models enabled us to conduct a robust evaluation of the RWKVTTS system across diverse linguistic contexts.

Additionally, we acknowledge the computational resources provided by yuanshi co., which were instrumental in training and evaluating the RWKV-7 model. This work was supported by RWKV Community, whose financial assistance made this research possible.

Finally, we extend our appreciation to the anonymous reviewers for their constructive comments and suggestions, which greatly improved the quality of this paper. Any remaining errors or shortcomings are our own.

\bibliographystyle{named}
\bibliography{ijcai25}

\end{document}